\documentclass[12pt]{article}

\usepackage{natbib}
\usepackage[dvips]{graphicx}
\usepackage{verbatim}

\usepackage{amssymb}
\usepackage{amsmath}
\usepackage{url}

\begin{document}

\title{Principal component proxy tracer analysis}

\author{\textbf{Peter Mills}\\
\textit{peteymills@hotmail.com}}

\maketitle

%\begin{center}
%{\LARGE {PRINCIPAL COMPONENT PROXY
%\vspace{0.2cm}
%TRACER ANALYSIS}}

\vspace{0.5cm}

%{\Large Peter Mills\\}
%\vspace{0.2cm}
%Peteysoft Foundation\\
%1159 Meadowlane\\Cumberland ON\\ K4C 1C3\\ Canada\\\textit{petey@peteysoft.org}
%\end{center}

%\section*{Abstract}
\begin{abstract}
We introduce a powerful method for dynamical reconstruction of long-lived tracers such as ozone.  It works by correlating the principal components of a matrix representation of the tracer dynamics with a series of sparse measurements. The method is tested on the 500 K isentropic surface using a simulated tracer and with ozone measurements from the Polar Aerosol and Ozone Measurement (POAM) III satellite instrument. The Lyapunov spectrum is measured and used to quantify the lifetime of each principal component. Using a 60 day lead time and five (5) principal components, cross validation of the reconstructed ozone and comparison with ozone sondes return root-mean-square errors of 0.20 ppmv and 0.47 ppmv, respectively.

\end{abstract}

\subsection*{Keywords}
\textbf{tracer dynamics,
interpolation,
inverse methods,
remote sensing,
chaos in the atmosphere,
data assimilation,
ozone}

%\linenumbers

%% main text

\newcommand{\vect}[1]{\ensuremath{\boldsymbol #1}}

\section{Introduction}

\citet{Randall_etal2002} demonstrate a method for proxy tracer reconstruction 
of ozone that works by correlating sparse measurements from satellites or
other remote-sensing instruments with conserved 
tracer fields such as potential vorticity.
The proxy tracer method is an approximate interpolation method, appropriate for
 long-lived tracers such as ozone, that takes into account the wind
dynamics.
Here we demonstrate a similar method in which the tracer dynamics are
represented as a matrix.
The matrix is decomposed using principal component analysis (PCA),
also called singular value decomposition (SVD), and the largest 
principal components then fitted to the measurements.

\section{Background}

Consider a system of ordinary differential equations:
\begin{equation}
\frac{\mathrm d \vect x}{\mathrm d t} = \vect f(\vect x,~t)
\end{equation}
We linearize this about $\vect x$:
\begin{eqnarray}
\frac{\mathrm d}{\mathrm d t} (\vect x + \delta \vect x) & \approx & \vect f + 
		\nabla \vect f \cdot \delta \vect x \\
\frac{\mathrm d}{\mathrm d t} \delta \vect x & \approx & \nabla \vect f \cdot \vect x
\end{eqnarray}
Define $H$ such that:
\begin{eqnarray} 
\frac{\mathrm d}{\mathrm d t} H & = & \nabla \vect f \cdot H \\
H (t=0) & = & I
\end{eqnarray}
where $I$ is the identity matrix.
This is what is known as the \textit{tangent model} which we 
decompose using principal component analysis (PCA), also called singular
value decomposition (SVD) or empirical orthogonal function (EOF) analysis:
\begin{equation}
H  =  U \cdot S \cdot V^T
\end{equation}
where $U$ and $V$ are orthonormal matrices while $S$ is diagonal and contains
the \textit{singular values} \citep{Press_etal1992}.
The matrix $U$ contains the \textit{left singular vectors} 
while $V$ contains the \textit{right singular vectors}.
The terms principal component (PC) and singular vector will be used
interchangeable to denote a left singular vector.

Singular vectors are increasingly being used in meteorology to quantify the
predictability of a forecast or to generate perturbations for ensemble
forecasts \citep{Tang_etal2006}.  
An Eulerian tracer simulation is linear, that is
the tangent vector is simply the dynamics.  Thus:
\begin{equation}
\vect x_j \approx H_j \cdot \vect x_0
\end{equation}
where $\vect x_j=\vect x(t_j)$ is a gridded representation of a 
passive flow tracer at the $j$th time step.
$H_j=A_j \cdot A_{j-1} \cdot A_{j-2} \cdot ~ ... ~ A_3 \cdot A_2 \cdot A_1$
is the tracer dynamics and the matrix $A_j$ maps $x_{j-1}$ to $x_j$.
We can say that:
\begin{equation}
A_j = \int_{t_{j-1}}^{t_j} \nabla \vect f \mathrm d t
\end{equation}

The Lyapunov exponents are defined as the logarithms of the time averages
of the singular values in the limit as time goes to infinity:
\begin{equation}
\lambda_i = \lim_{t \rightarrow \infty} \frac{1}{t} \log s_i;
~~~~~~~\lambda_{i-1} \le \lambda_i
\end{equation}
where $s_i$ is the $i$th singular value \citep{Ott1993}.
For most systems:
\begin{equation}
|\delta \vect x| \approx |\delta \vect x(0) | \exp(\lambda_i)
\label{lambda1}
\end{equation}
That is, as $H$ is integrated forward, the largest singular value and
the largest singular vector will increasingly begin to dominate
\citep{Ott1993}.

\section{Numerics and data}

To run the tracer advection, the \textit{ctraj} software package is used
(http://ctraj.sf.net).  The codes are written in C++ and contain programs
for gridded, semi-Lagrangian tracer advection on an 
azimuthally-equidistant-projected coordinate system.
Two fields are advected simultaneously, one for the Northern hemisphere
and one for the Southern hemisphere, with equatorial crossings accounted for.
Gridding on both hemispheres is 100 by 100, or 200km-,
1.8 degree-latitude-separation at the pole.  
Output is written to a series of sparse matrices which are then
decomposed with the Lanczos method \citep{Golub_Van_Loan1996} using the
Arnoldi package (ARPACK) \citep{Lehoucq_Scott1996}.

The Polar Ozone and Aerosol Measurement (POAM) III instrument is a solar-
occultation instrument mounted on a sun-synchronous, low-earth-orbit
satellite \citep{Lucke_etal1999}.  
Using optimal estimation \citep{Rodgers2000}, ozone profiles are retrieved 
within a narrow latitude band in either polar region \citep{Lumpe_etal2002}.  
It is capable of returning 28 or 29 measurements per day,
alternating between Northern and Southern hemisphere, however because of
a malfunction
in the instrument, it normally operates in only one or the other hemisphere for longer periods.  Therefore, we confine ourselves to earlier data,
October and November 1998, when more frequent and diverse measurements
are available.

The National Center for Environmental Prediction (NCEP) supplies, 
free-of-charge,
gridded (2.5 by 2.5 degrees longitude/latitude, 4 time daily), reanalyzed 
climate data starting in 1948 \citep{Kalnay_etal1996}.
Wind and temperature data is used to drive the advection model.

\section{Tracer correlation}

\begin{figure}
\begin{center}
\includegraphics[angle=90,width=0.9\textwidth]{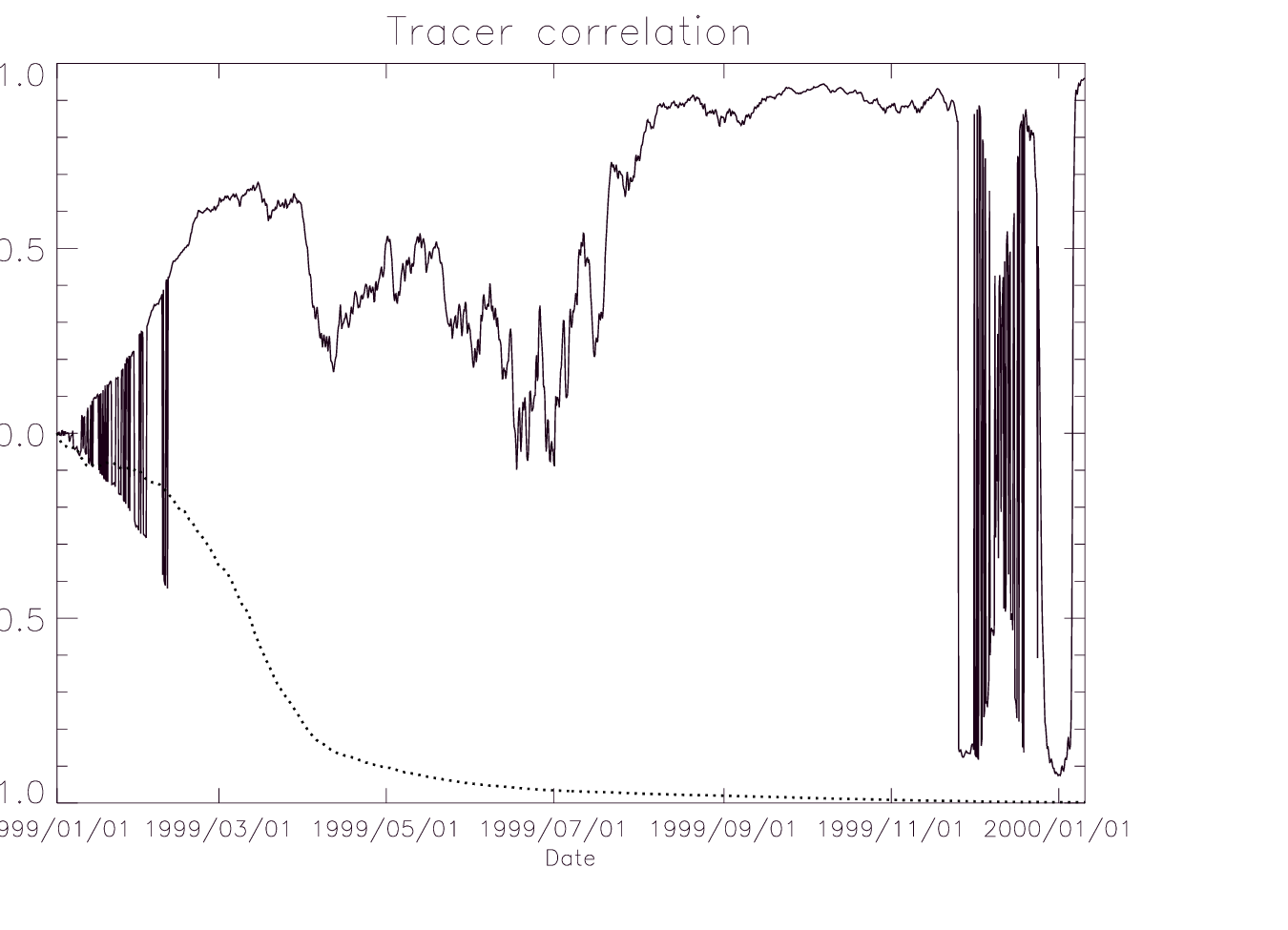}
\caption{The correlation over time of two differently-initialized tracers
(broken line)--zonally symmetric and meridionally symmetric--and of
the zonally-symmetric-initialized tracer with the first principal component.
The simulation was driven with NCEP reanalysis 1 data on the 500 K isentropic
level with an Eulerian time-step of six (6) hours and a Lagrangian time-step
of one (1) hour.}\label{tcorr}
\end{center}
\end{figure}

\begin{figure}
\includegraphics[width=0.45\textwidth]{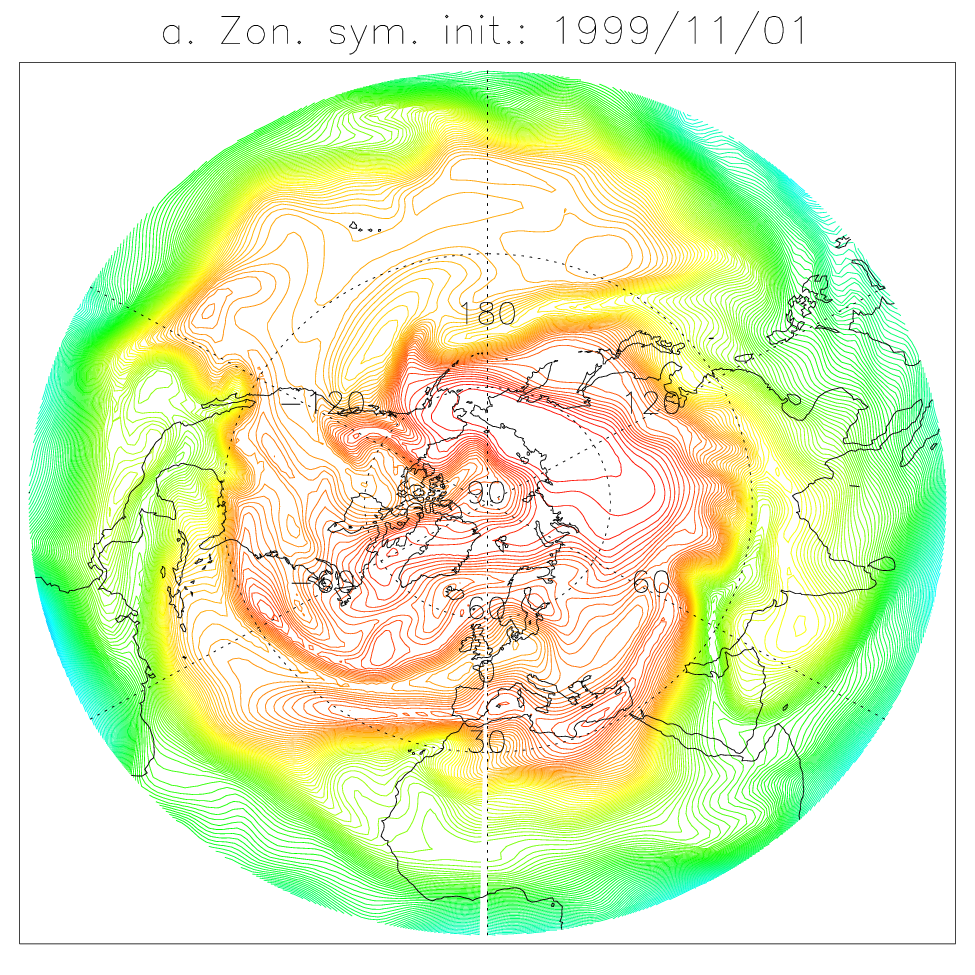}
\includegraphics[width=0.45\textwidth]{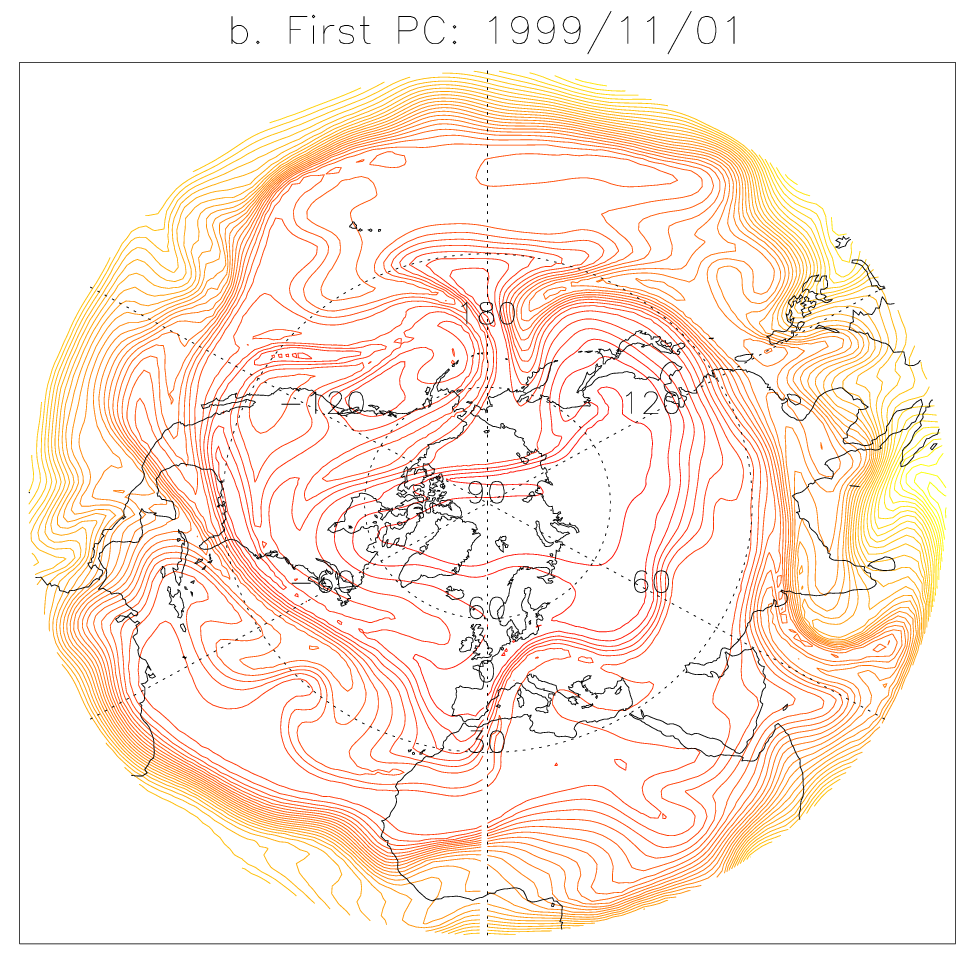}
\caption{Comparison of a simulated tracer (a.) and the first principal
component (b.) for the same time integration.}\label{pc1}
\end{figure}

Two differently-initialized tracers, when integrated with the same
wind fields over a long time period, become correlated.
This can be used to infer global fields of a long-lived tracer such as
ozone based on only a few sparse measurements 
\citep{Allen_Nakamura2003,Randall_etal2002}.
Figure \ref{tcorr} demonstrates this with the extreme example of an initially
zonally-symmetric tracer and an initialy meridionally-symmetric tracer.
Tracers are advected with National Center for Environmental Prediction
(NCEP) reanalysis 1 data at the 500 K isentrop \citep{Kalnay_etal1996}.

We also plot the correlation of the first tracer with the largest singular
vector.  We see that, because of Equation (\ref{lambda1}), they too become
correlated over time.
This at least partially explains the efficacy of the proxy tracer method.
A sample PC as compared with the tracer is shown in Figure \ref{pc1}.  

\begin{figure}
\begin{center}
\includegraphics[angle=90,width=0.9\textwidth]{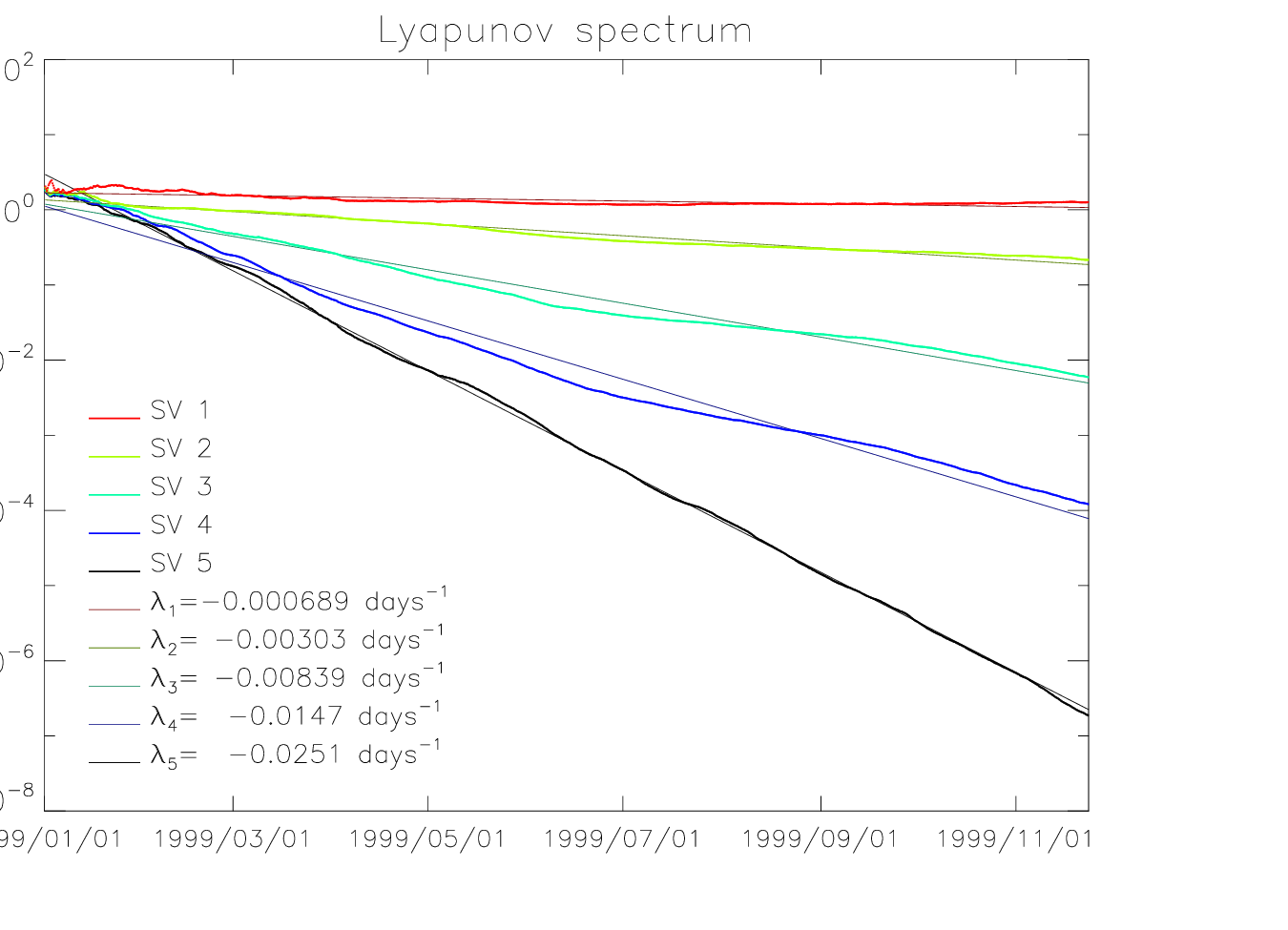}
\caption{Plot of the top five (5) singular-values of a semi-Lagrangian
tracer simulation over time.  Straight-line fits return the Lyapunov-spectrum.
The simulation was done on the 500 K isentropic level.}
\label{lyap_spec}
\end{center}
\end{figure}

Figure \ref{lyap_spec} plots the time evolution of the singular values.
From this we can calculate the Lyapunov spectrum by making straight line
fits of their logarithms.
While the resulting fields may develop into complex fractals \citep{Mills2009}
%Raise my h-index to 2.  Yay!
the Lyapunov spectrum shows that the tracer dynamics themselves 
are not truly chaotic, but are
only on the cusp: the largest singular value remains approximately constant.
It also shows how quickly the other singular vectors decay,
so that the largest will eventually dominate in accordance with
Equation \ref{lambda1}.

\section{Principal component proxy}
\label{pc_proxy}

\begin{figure}
\begin{center}
\includegraphics[width=1\textwidth]{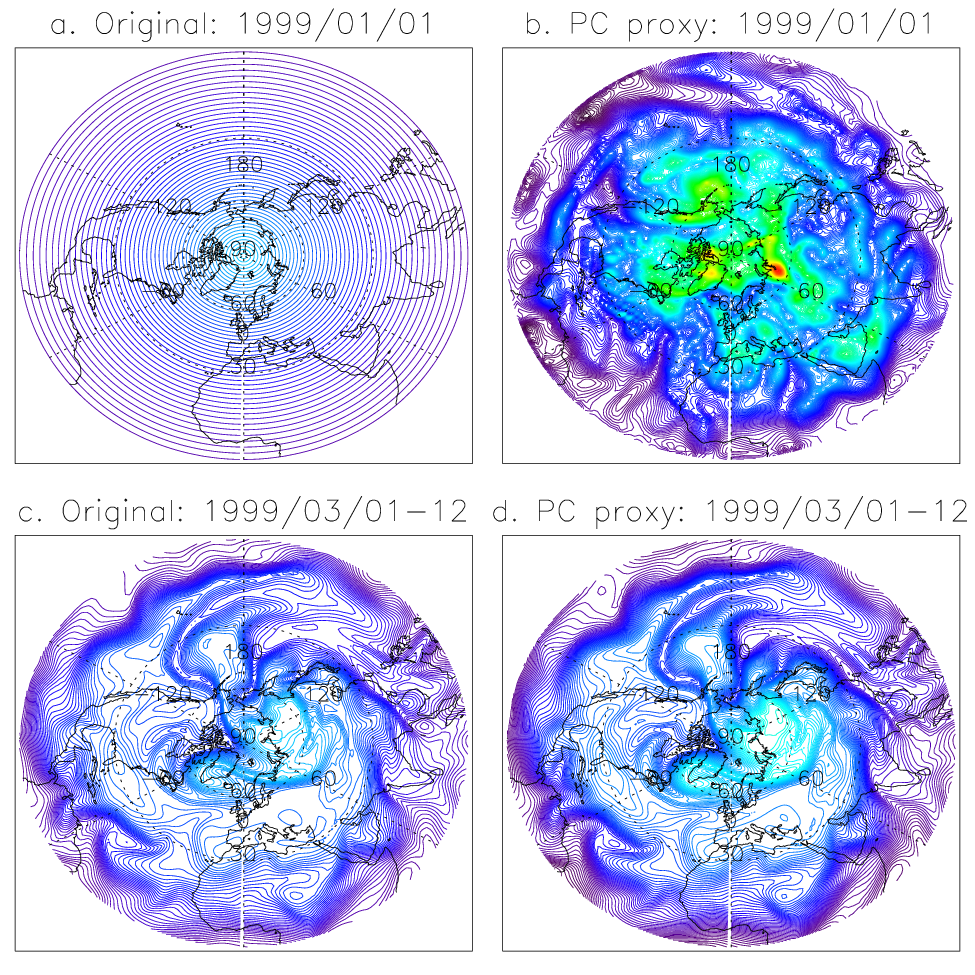}
\caption{Comparison of simulated tracer fields with the reconstructed version.
a. and b. compare the initial field of the simulated versus the reconstructed,
respectively, while c. and d. compare the fields at the lead time,
60 days later.}
\label{proxyfields}
\end{center}
\end{figure}

\begin{figure}
\begin{center}
\includegraphics[angle=90, width=0.9\textwidth]{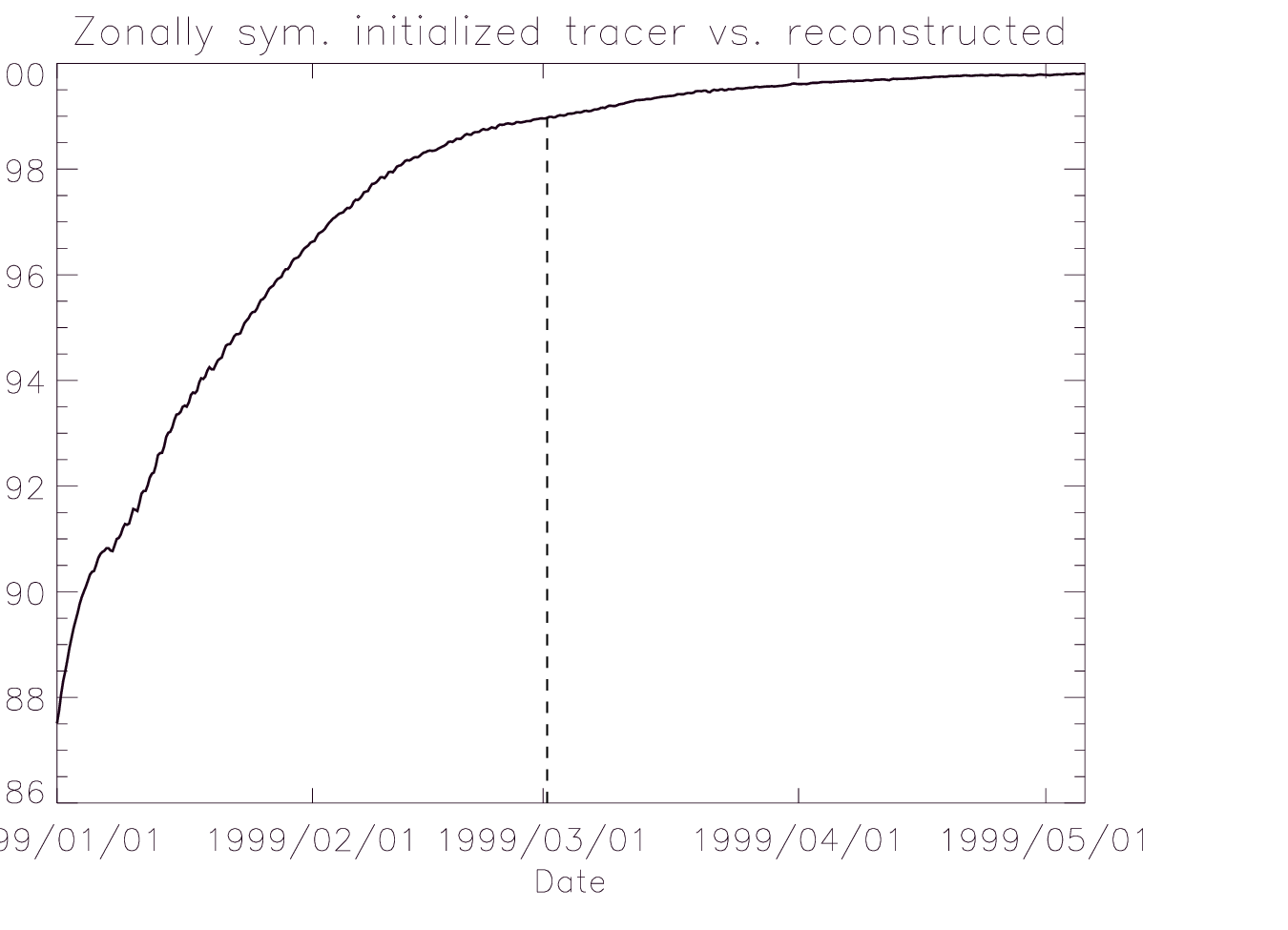}
\caption{Correlation over time of a zonally-symmetric-initialized, passive tracer
with one reconstructed using PC proxy using a lead time of 60 days,
marked by the vertical, dashed line.}\label{PCproxytest}
\end{center}
\end{figure}

Principal component proxy analysis works by linearly regressing the
largest singular vectors with the measurements.
Since the measurements will likely not occur all at the same time,
we must first generate the right singular vectors for a given tracer 
mapping, $H_j$, at \textit{lead time} $j$, then generate a series of
left singular vectors for different time steps
by applying the tracer mapping.
For the right singular vectors, ARPACK is used to compute the
eigenvectors of $H_j^T\cdot H_j$.
Values interpolated within the right singular vectors at the 
measurement locations are fitted
through coefficients, $\lbrace c_i \rbrace$, to the measurements.
Thus, we need to fit the following equation:
\begin{equation}
\sum_{i=1}^m c_i \vect w_k \cdot H_k \cdot \vect v_i = q_k
\end{equation}
where $m$ is the number of singular vectors used in the analysis,
$q_k$ is the measurement at time step $k$ (we assume that each measurement
has a unique time stamp), $H_k$ is the tracer mapping at time $k$,
$w_k$ is a vector of interpolation coefficients and $\vect v_i$ is
the $i$th right principal component.
The fitting is done using a linear least squares \citep{gsl_ref}.

To perform the analysis, we need to choose a lead time, as well as
a \textit{measurement window}.
The lead time determines how long the tracer is advected before performing
the SVD.
Measurements are selected within the measurement window which is centered
at the end of the lead time.

\begin{figure}
\begin{center}
\includegraphics[angle=90,width=0.8\textwidth]{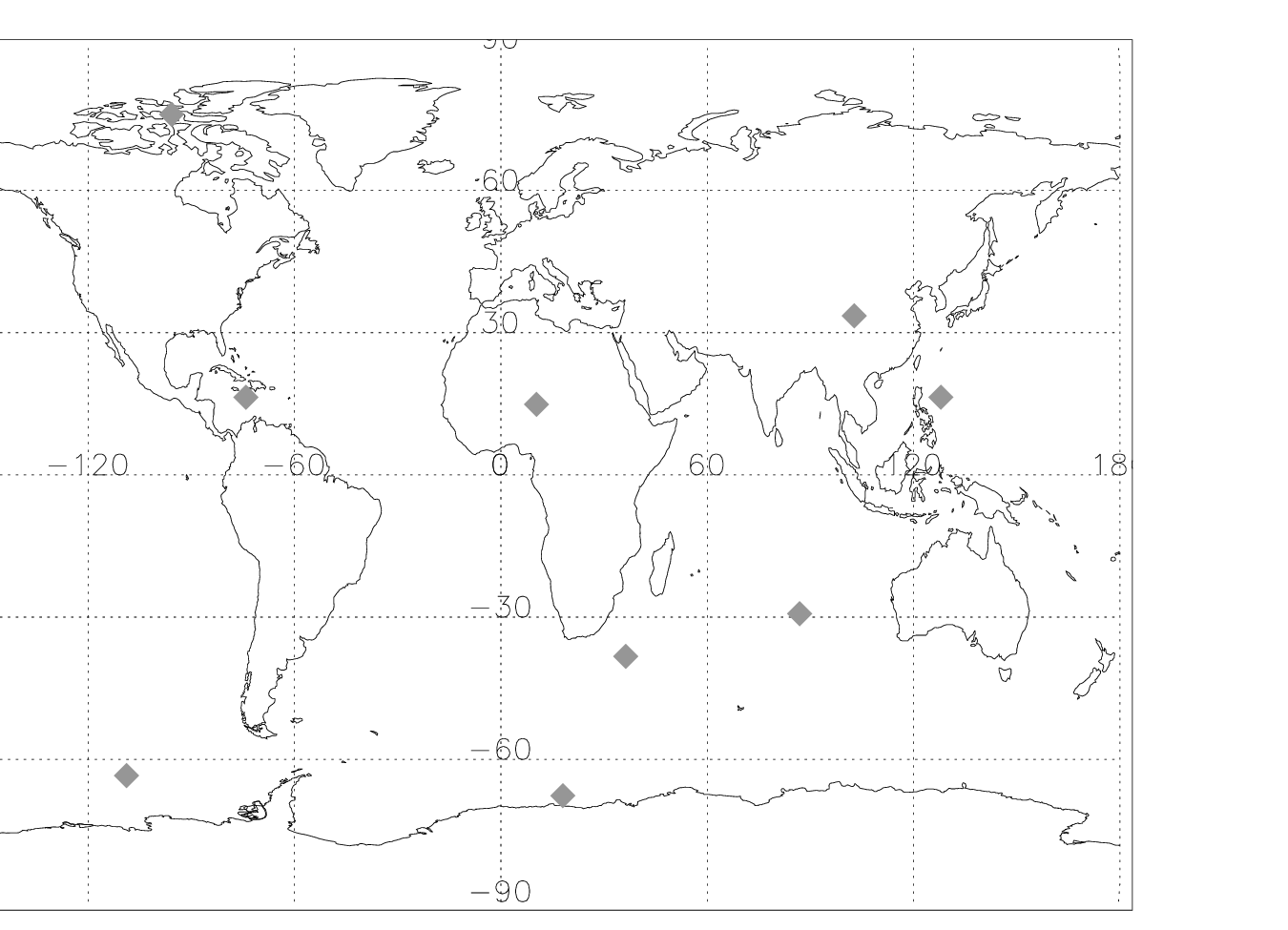}
\caption{Locations of the ten (10) simulated sparse measurements used for the
test retrieval.}\label{sparse}
\end{center}
\end{figure}

Figure \ref{proxyfields} and \ref{PCproxytest} 
shows the results of a test retrieval for a 
zonally-symmetric-initialized tracer using a lead time of 60 days
and five singular vectors.
The Lyapunov spectrum can help us select the number of singular vectors
as it shows how many remain significant at
a given lead time--see Figure \ref{lyap_spec}.
Ten sparse measurements were randomly selected in space and time
within a measurement window of one day--these are plotted in Figure \ref{sparse}.
The Pearson correlation for the initial field (Figures \ref{proxyfields}a. and b.)
is 0.875, while the correlation at the lead time 
(Figures \ref{proxyfields}c. and d.) is 0.99

\section{Ozone reconstruction}

\begin{figure}
\begin{center}
\includegraphics[width=1\textwidth]{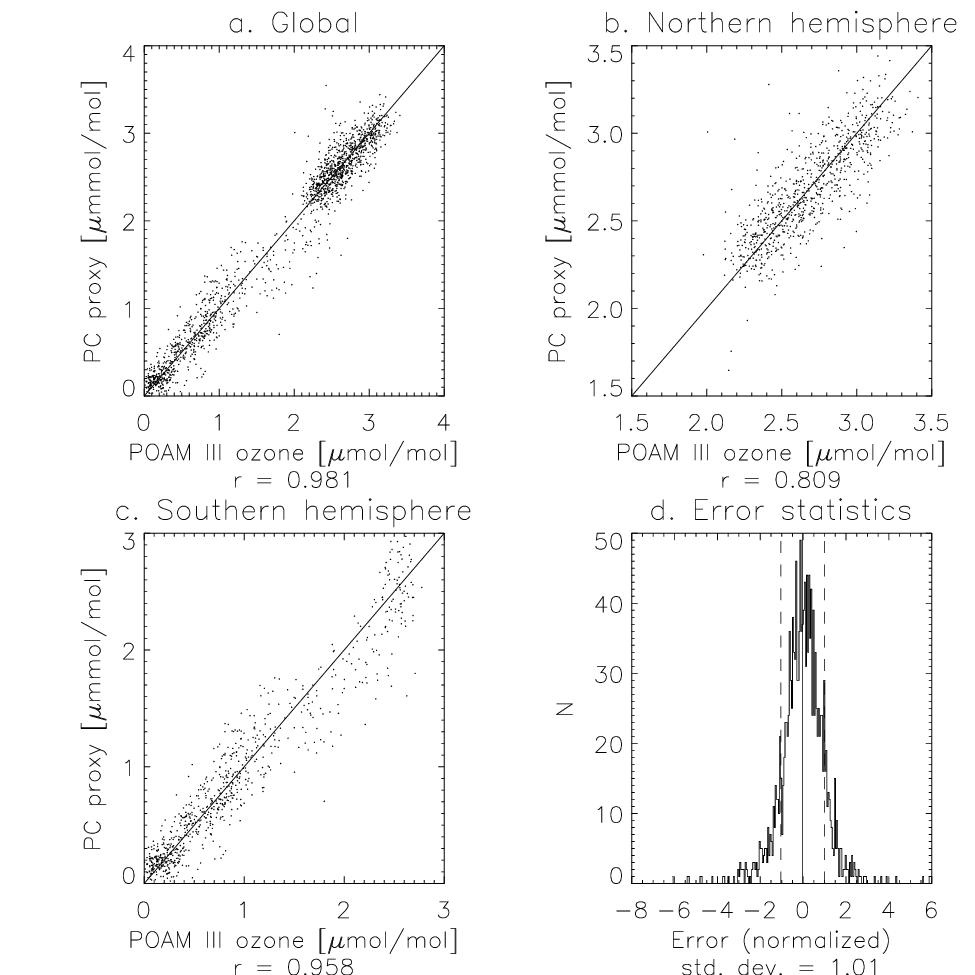}
\caption{a. Scatter plot of PC proxy cross-validation results
with POAM III ozone data on the 500 K isentrop for November
and December 1998, 60 day lead time and 2 day measurement window.
b. and c. show comparison results restricted to Northern and Southern hemisphere
respectively.  c. Histogram of errors normalized by original error
estimates from the POAM inversion.
The solid vertical lines shows the average, while the dashed lines show the standard
deviation.}\label{POAM_proxyscat}
\end{center}
\end{figure}

The purpose here is not to perform a rigorous validation, but rather
to demonstrate \textit{proof-of-concept}.
To do this we perform a cross-validation exercise on Polar Ozone
and Aerosol Measurement (POAM) III satellite retrievals \citep{Lucke_etal1999}
in which the data is
divided into two parts: a \textit{training} data set and 
a \textit{test} data set.  The two are divided approximately evenly:
each measurement is selected at random to go into one or the other set.
The analysis is done once again at the 500 K isentropic level; 
a lead time of 60 days and a measurement window of 2 days is used.

Validation results are shown in Figure \ref{POAM_proxyscat}.
Unlike in \citet{Randall_etal2002}, the reconstruction is done over
the entire globe.
Since ozone concentrations are generally lower in the Southern
hemisphere, this will produce an artificially high skill score,
thus we also include comparisons limited to each of the
two hemispheres.
Figure \ref{POAM_proxyscat}d. shows the error statistics, 
normalized by the original error from the POAM inversion \citep{Lumpe_etal2002}.
Note that the standard deviation for this error is almost exactly
one (1), meaning that the error for the reconstructed ozone is
about the same as the original measurement error.

\begin{figure}
\begin{center}
\includegraphics[width=1\textwidth]{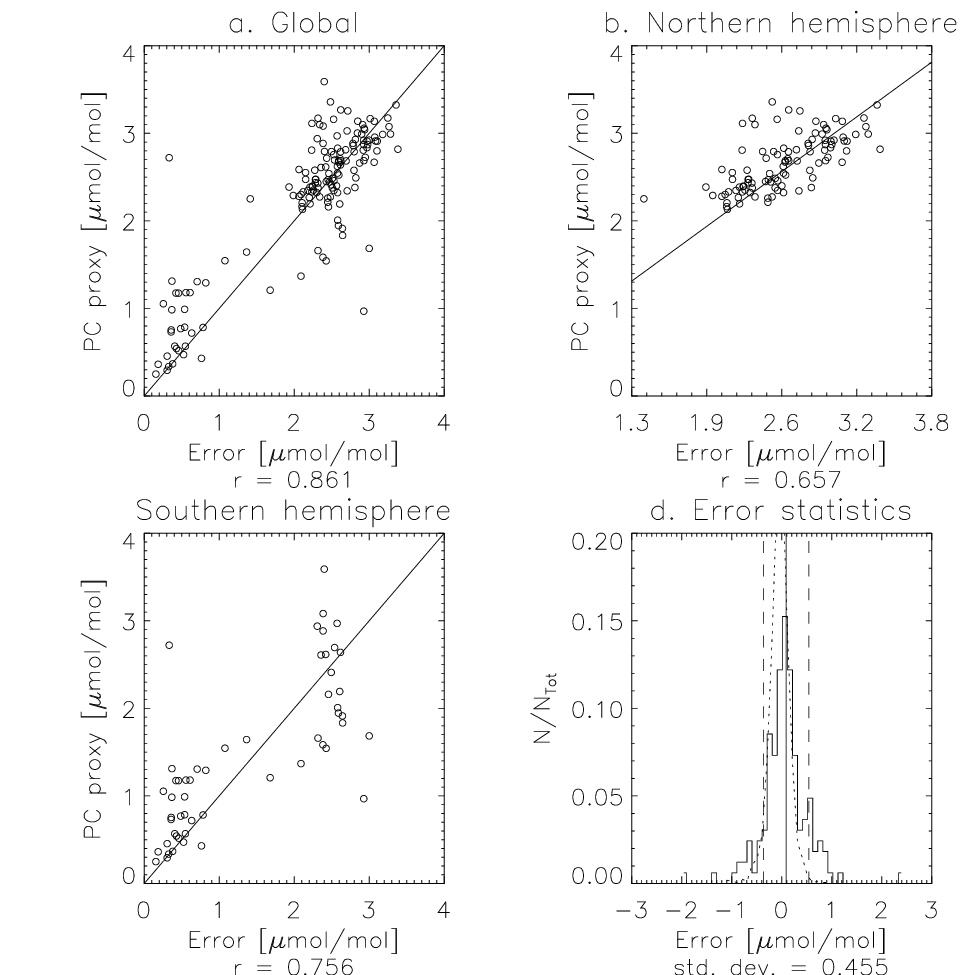}
\caption{Ozone reconstructed from POAM versus ozone sonde data.
Figures b. and c. restrict the comparison to the Northern and Southern
hemispheres respectively.
Figure d. shows error statistics (histogram bars).
Solid vertical line is the average, while the dashed lines show
the standard deviation.  Error statistics for the cross-validation
are shown for comparison (dotted line).}
\label{proxyvssonde}
\end{center}
\end{figure}

\begin{figure}
\begin{flushright}
\includegraphics[angle=90,width=0.9\textwidth]{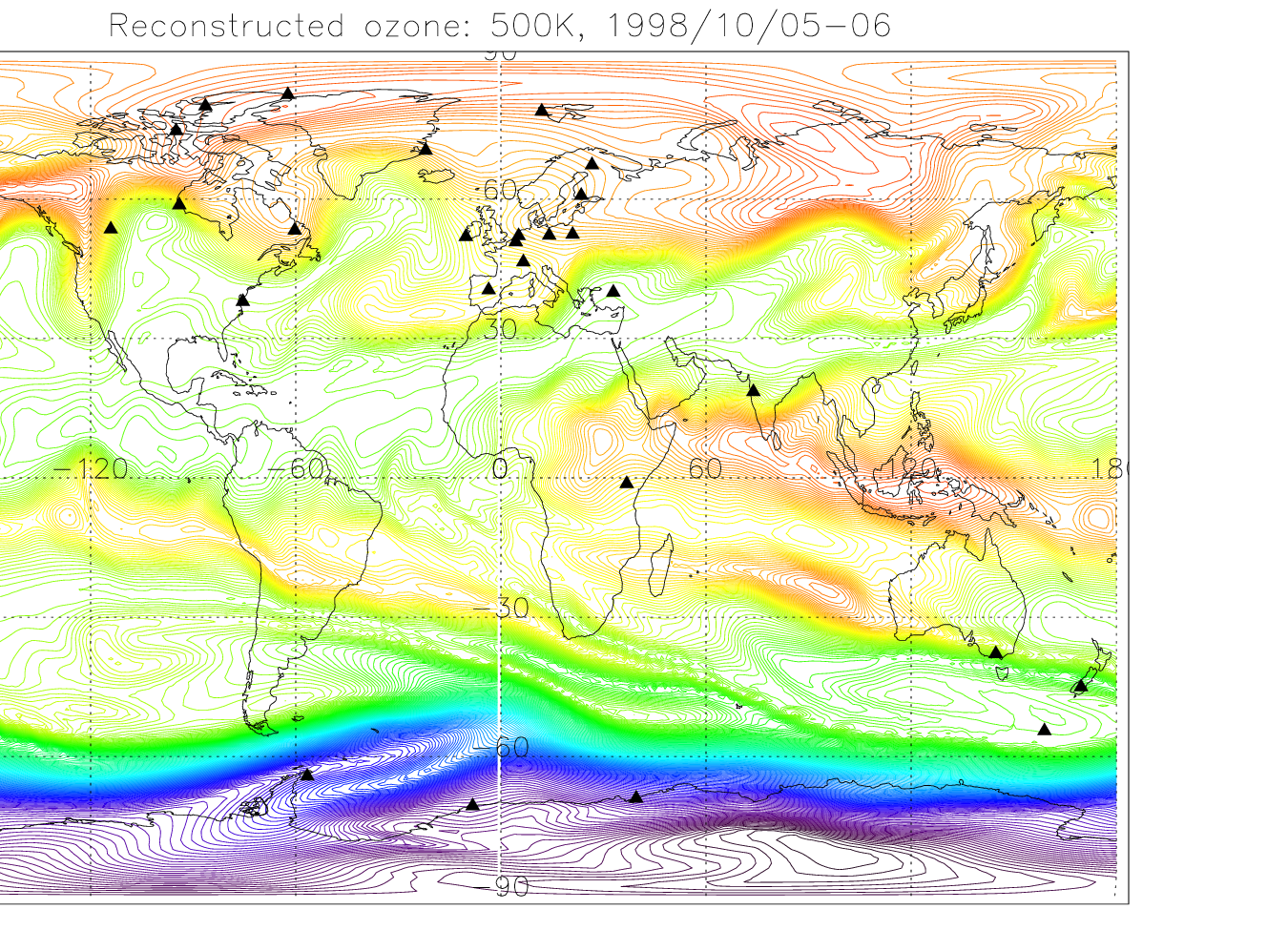}
\end{flushright}
\begin{flushleft}
\includegraphics[width=.9\textwidth]{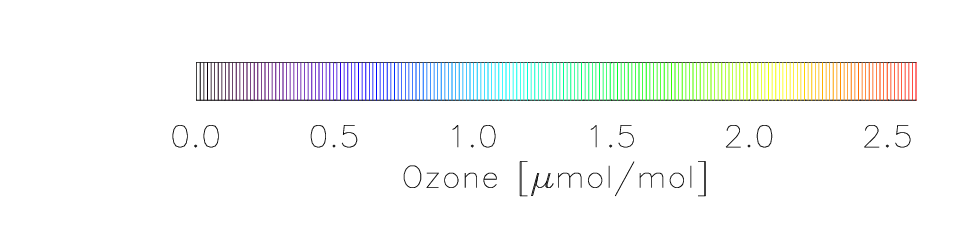}
\end{flushleft}
\caption{Sample ozone field reconstructed using the PC proxy technique
from POAM III data at the 500 K isentrop.
Triangles show the locations of the ozone sonde launch stations
used in the validation exercise.}\label{sample_ozone}
\end{figure}

Finally, we compare the ozone fields reconstructed from the POAM data 
with ozone sonde measurements from the World Ozone Data Centre (WODC)
\citep{Hare_etal2000}.
Validation results are shown in Figure \ref{proxyvssonde}.
Once again, the comparison is done for both Northern and
Southern hemispheres exclusively, using the globally reconstructed
ozone, and error statistics are shown in the final figure, 
\ref{proxyvssonde}d.
Root-mean-square error for the ozone sondes is 0.45 $\mu$mol/mol
(volume-mixing-ratio (vmr) in parts-per-million (ppm): ppmv) while
for the cross-validation (statistics are graphed with the dotted line) it is 
0.20 $\mu$mol/mol.

A sample reconstructed ozone field is shown in Figure \ref{sample_ozone}, 
Note that for the sample field, values towards the equator and the
lower latitudes are suspect, discussed in more detail in 
Section \ref{discussion}, below.
The launch stations are plotted on the field and, like the POAM
measurements, are mostly restricted to the higher and mid-latitudes.

\section{Discussion and conclusions}

\label{discussion}
Further work needs to be done to determine the optimal number
of PCs to use in an analysis as well as the
optimal lead times.
Preliminary work shows that beyond a lead time of about 60 days,
skill scores differ little.
Presumably, reconstruction of shorter-(longer-)lived tracers would work better
with shorter (longer) lead times.
Naturally, a longer lead time means longer compute times.
The number of PCs required will
be related in part to the lead time, as discussed in Section \ref{pc_proxy},
with shorter lead times requiring more PCs.
Obviously, a larger number of measurements will support the use of
more PCs.

Another problem in some of the reconstructed fields is that the solution
tended to blow up towards the equator, showing ringing, negative concentrations,
and other artifacts.
Including a constant term in the fitting procedure did little 
to alleviate the problem.
A better solution would be to use some form of regularization to constrain
the solution, as in optimal estimation \citep{Rodgers2000}, or better still,
include more measurements in the analysis.
Most of the ozone sondes were also launched in the higher
latitudes, so this did not significantly affect the comparison results.

Nonetheless, principal-component (PC) proxy tracer analysis
is shown to be a powerful, dynamical interpolation method
capable of reconstructing a passive tracer using as few as 
ten (10) sparse measurements.
Ozone reconstructed from the Polar Ozone and Aerosol (POAM) III
instrument showed reasonable agreement with ozone sondes,
despite using, on average, 52 measurements per field and these
limited to a narrow latitude bands in either hemisphere.

The method has the advantage over more tradition proxy tracer analysis
in that it provides more degrees of freedom from the higher principal 
components thus can account for more recent changes in the tracer.

The ozone reconstruction for the cross-validation exercise was so good,
in fact, that the errors were of the same order as the original POAM
retrievals, suggesting that these error estimates are too high to
begin with.

Software for this project can be found at: \url{http://ctraj.sf.net}.

\section*{Acknowledgements}
Thanks to the National Center for Environmental Prediction (NCEP) for
reanalysis data and to my former colleagues at the 
Naval Research Laboratory for POAM data.
Thanks also to the World Ozone Data Centre (WODC) for radiosonde ozone
data.

The author would also like to thank John Lin of the University of Waterloo
for partially supporting this work and the Institute of Environmental Physics,
University of Bremen for the use of their computers.

\bibliographystyle{apa}
\bibliography{PC_proxy}

\begin{thebibliography}{}

\bibitem[\protect\astroncite{Allen and Nakamura}{2003}]{Allen_Nakamura2003}
Allen, D.~R. and Nakamura, N. (2003).
\newblock Tracer {E}quivalent {L}atitude: {A} {D}iagnostic {T}ool for
  {I}sentropic {T}ransport {S}tudies.
\newblock {\em Journal of the Atmospheric Sciences}, 60:287--303.

\bibitem[\protect\astroncite{Galassi et~al.}{2007}]{gsl_ref}
Galassi, M., Davies, J., Theiler, J., B.~Gough, B., Jungman, G., Booth, M., and
  Ross, F. (2007).
\newblock {\em {GNU} {S}cientific {L}ibrary: Reference Manual}.
\newblock Available online at: \url{http://www.gnu.org/software/gsl.}

\bibitem[\protect\astroncite{Golub and van Loan}{1996}]{Golub_Van_Loan1996}
Golub, G.~H. and van Loan, C.~F. (1996).
\newblock {\em Matrix Computations}.
\newblock Johns Hopkins University Press.

\bibitem[\protect\astroncite{Hare et~al.}{2000}]{Hare_etal2000}
Hare, E.~W., Carty, E.~J., and Wardle, D.~I. (2000).
\newblock Guide to the {WMO}/{GAW} world ozone data centre.
\newblock Technical report, Meteorological Service of Canada, Environment
  Canada.

\bibitem[\protect\astroncite{Kalnay et~al.}{1996}]{Kalnay_etal1996}
Kalnay, E., Kanamitsu, M., Kistler, R., Collins, W., Deaven, D., Gandin, L.,
  Iredell, M., Saha, S., White, G., Woollen, J., Zhu, Y., Chelliah, M.,
  Ebisuzaki, W., Higgins, W., Janowiak, J., Mo, K., Ropelewski, C., Wang, J.,
  Leetmaa, A., Reynolds, R., Jenne, R., and Joseph, D. (1996).
\newblock The {NCEP}/{NCAR} 40-year reanalysis project.
\newblock {\em Bull. Amer. Meteor. Soc.}, 77:437--470.

\bibitem[\protect\astroncite{Lehoucq and Scott}{1996}]{Lehoucq_Scott1996}
Lehoucq, R.~B. and Scott, J.~A. (1996).
\newblock An {E}valuation of {S}oftware for {C}omputing {E}igenvalues of
  {S}parse {N}onsymmetric {M}atrices.
\newblock Technical Report MCS-P547-1195, Argonne National Laboratory.

\bibitem[\protect\astroncite{Lucke et~al.}{1999}]{Lucke_etal1999}
Lucke, R.~L., Korwan, D.~R., Bevilacqua, R.~M., Hornstein, J.~S., Shettle,
  E.~P., Chen, D.~T., Daehler, M., Lumpe, J.~D., Fromm, M.~D., Debrestian, D.,
  Neff, B., Squire, M., König-Langlo, G., and J.~Davies, J. (1999).
\newblock The {P}olar {O}zone and {A}erosol {M}easurement({POAM}) {III}
  instrument and early validation results.
\newblock {\em Journal of Geophysical Research}, 104(D15):18785--18799.

\bibitem[\protect\astroncite{Lumpe et~al.}{2002}]{Lumpe_etal2002}
Lumpe, J.~D., Bevilacqua, R.~M., Hoppel, K.~W., and Randall, C.~E. (2002).
\newblock {POAM} {III} retrieval algorithm and error analysis.
\newblock {\em Journal of Geophysical Research}, 107(D21):ACH5.1--ACH5.32.

\bibitem[\protect\astroncite{Mills}{2009}]{Mills2009}
Mills, P. (2009).
\newblock Isoline retrieval: {A}n optimal method for validation of advected
  contours.
\newblock {\em Computers \& Geosciences}, 35(20):2020--2031.

\bibitem[\protect\astroncite{Ott}{1993}]{Ott1993}
Ott, E. (1993).
\newblock {\em Chaos in Dynamical Systems}.
\newblock Cambridge University Press.

\bibitem[\protect\astroncite{Press et~al.}{1992}]{Press_etal1992}
Press, W.~H., Teukolsky, S.~A., Vetterling, W.~T., and Flannery, B.~P. (1992).
\newblock {\em Numerical Recipes in C}.
\newblock Cambridge University Press, 2nd edition.

\bibitem[\protect\astroncite{Randall et~al.}{2002}]{Randall_etal2002}
Randall, C.~E., Lumpe, J.~D., Bevilacqua, R.~M., Hoppel, K.~W., Fromm, M.~D.,
  Salawitch, R.~J., Swartz, W.~H., Lloyd, S.~A., Kyro, E., von~der Gathen, P.,
  Claude, H., Davies, J., DeBacker, H., Dier, H., Molyneux, M.~J., and Sanchoi,
  J. (2002).
\newblock Reconstruction of three-dimensional ozone fields using {POAM} {III}
  during {SOLVE}.
\newblock {\em Journal of Geophysical Research}, 107(D20):8299--8312.

\bibitem[\protect\astroncite{Rodgers}{2000}]{Rodgers2000}
Rodgers, C.~D. (2000).
\newblock {\em Inverse Methods for Atmospheric Sounding: Theory and Practice}.
\newblock World Scientific.

\bibitem[\protect\astroncite{Tang et~al.}{2006}]{Tang_etal2006}
Tang, Y.~R., Kleeman, R., and Miller, S. (2006).
\newblock {ENSO} predictability of a {F}ully {C}oupled {GCM} {M}odel {U}sing
  {S}ingular {V}ector {A}nalysis.
\newblock {\em Journal of Climate}, 19(14):3361--3377.

\end{thebibliography}

\end{document}